\documentclass[%
reprint,
superscriptaddress,
 amsmath,amssymb,
 aps,
 prx,
floatfix,
]{revtex4-1}

\usepackage[utf8]{inputenc}
\usepackage{xcolor}
\usepackage[margin=0.75in]{geometry}
\usepackage{graphicx} 
\usepackage{physics}
\usepackage{amsmath}
\usepackage{siunitx}

\newcommand{\sixj}[6]{ \begin{Bmatrix}
  #1 & #2 & #3 \\
  #4 & #5 & #6 
 \end{Bmatrix}}

\usepackage[draft]{changes}
\begin{document}


\title{Erasure conversion for fault-tolerant quantum computing in alkaline earth Rydberg atom arrays}

\author{Yue Wu}
\affiliation{Yale University, Department of Computer Science, New Haven, CT 06520}
\author{Shimon Kolkowitz}
\affiliation{University of Wisconsin-Madison, Department of Physics, Madison, WI 53706}
\author{Shruti Puri}
\affiliation{Yale University, Department of Applied Physics, New Haven, CT 06520}
\author{Jeff D. Thompson}
\email{jdthompson@princeton.edu}
\affiliation{Princeton University, Department of Electrical and Computer Engineering, Princeton, NJ, 08544}
\date{\today}

\begin{abstract}
Executing quantum algorithms on error-corrected logical qubits is a critical step for scalable quantum computing, but the requisite numbers of qubits and physical error rates are demanding for current experimental hardware. Recently, the development of error correcting codes tailored to particular physical noise models has helped relax these requirements. In this work, we propose a qubit encoding and gate protocol for ${}^{171}$Yb neutral atom qubits that converts the dominant physical errors into \emph{erasures}, that is, errors in known locations. The key idea is to encode qubits in a metastable electronic level, such that gate errors predominantly result in transitions to disjoint subspaces whose populations can be continuously monitored via fluorescence. We estimate that 98\% of errors can be converted into erasures. We quantify the benefit of this approach via circuit-level simulations of the surface code, finding a threshold increase from $0.937\%$ to $4.15\%$. We also observe a larger code distance near the threshold, leading to a faster decrease in the logical error rate for the same number of physical qubits, which is important for near-term implementations. Erasure conversion should benefit any error correcting code, and may also be applied to design new gates and encodings in other qubit platforms.
\end{abstract}

\maketitle

\section{Introduction}
Scalable, universal quantum computers have the potential to outperform classical computers for a range of tasks \cite{montanaro2016}. However, the inherent fragility of quantum states and the finite fidelity of physical qubit operations make errors unavoidable in any quantum computation. Quantum error correction \cite{shor1995,gottesman1997,knill1997} allows multiple physical qubits to represent a single logical qubit, such that the correct logical state can be recovered even in the presence of errors on the underlying physical qubits and gate operations. 

If the logical qubit operations are implemented in a fault-tolerant manner that prevents the proliferation of correlated errors, the logical error rate can be suppressed arbitrarily so long as the error probability during each operation is below a threshold \cite{aharonov2008,knill1996}. Fault-tolerant protocols for error correction and logical qubit manipulation have recently been experimentally demonstrated in several platforms \cite{egan2021,ryan-anderson2021,abobeih2021,postler2021}.

The threshold error rate depends on the choice of error correcting code and the nature of the noise in the physical qubit. While many codes have been studied in the context of the abstract model of depolarizing noise arising from the action of random Pauli operators on the qubit, the realistic error model for a given qubit platform is often more complex, which presents both opportunities and challenges. For example, qubits encoded in cat-codes in superconducting resonators can have strongly biased noise \cite{grimm2020}, leading to significantly higher thresholds \cite{aliferis2008,darmawan2021} given suitable bias-preserving gate operations for fault-tolerant syndrome extraction \cite{puri2020}. On the other hand, many qubits also exhibit some level of leakage outside of the computational space \cite{knill1996,preskill1997}, which requires extra gates in the form of leakage-reducing units, decreasing the threshold \cite{suchara2015}.

Another type of error is an erasure, or detectable leakage, which denotes an error at a known location. Erasures are significantly easier to correct than depolarizing errors in both classical \cite{coverthomasm2006} and quantum \cite{grassl1997, gottesman1997} settings. For example, a four-qubit quantum code is sufficient to correct a single erasure error \cite{grassl1997}, and the surface code threshold under the erasure channel approaches 50\% (with perfect syndrome measurements), saturating the bound imposed by the no-cloning theorem \cite{stace2009}. Erasure errors arise naturally in photonic qubits: if a qubit is encoded in the polarization, or path, of a single photon, then the absence of a photon detection signals an erasure, allowing efficient error correction for quantum communication \cite{muralidharan2014} and linear optics quantum computing \cite{knill2001,kok2007}. However, techniques for detecting the locations of errors in matter-based qubits have not been extensively studied.

\begin{figure}
    \centering
    \includegraphics[width=\columnwidth]{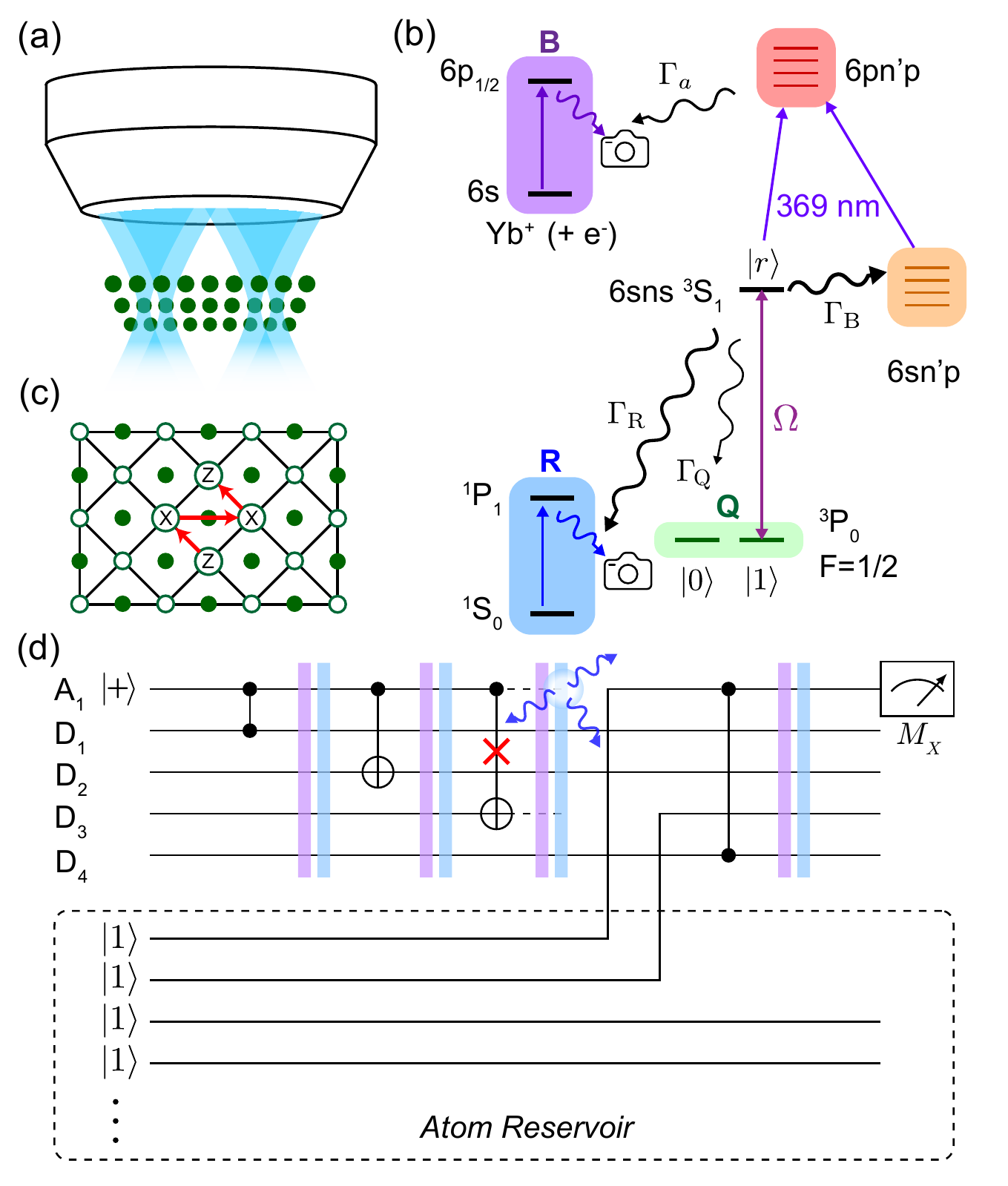}
    \caption{
    (a) Schematic of a neutral atom quantum computer.
    (b) The physical qubits are individual ${}^{171}$Yb atoms. The qubit states are encoded in the metastable $6s6p$ $^3$P$_0$ $F=1/2$ level (subspace Q), and two-qubit gates are performed via the Rydberg state $\ket{r}$, which is accessed through a single-photon transition ($\lambda = 302$ nm) with Rabi frequency $\Omega$. The dominant errors during gates are decays from $\ket{r}$ with a total rate $\Gamma = \Gamma_B + \Gamma_R + \Gamma_Q$. Only a small fraction $\Gamma_Q/\Gamma \approx 0.05$ return to the qubit subspace, while the remaining decays are either blackbody (BBR) transitions to nearby Rydberg states ($\Gamma_B/\Gamma \approx 0.61$) or radiative decay to the ground state $6s^2$ $^1$S$_0$ ($\Gamma_R/\Gamma \approx 0.34$). At the end of a gate, these events can be detected and converted into erasure errors by detecting fluorescence from ground state atoms (subspace R), or ionizing any remaining Rydberg population via autoionization, and collecting fluorescence on the Yb$^+$ transition (subspace B).
    (c) A patch of the XZZX surface code studied in this work, showing data qubits (open circles), ancilla qubits (filled circles) and stabilizer operations.
    (d) Quantum circuit representing a single stabilizer measurement in the XZZX surface code with erasure conversion. Erasure detection is applied after each gate, and erased atoms are replaced from a reservoir as needed using a moveable optical tweezer. }
    \label{fig:fig1}
\end{figure}

In this work, we present an approach to fault-tolerant quantum computing in Rydberg atom arrays \cite{jaksch2000,Lukin2001,Saffman2010} based on converting a large fraction of naturally occurring errors into erasures. Our work has two key components. First, we present a physical model of qubits encoded in a particular atomic species, ${}^{171}$Yb \cite{noguchi2011,ma2021,jenkins2021}, that enables erasure conversion without additional gates or ancilla qubits. By encoding qubits in the hyperfine states of a metastable electronic level, the vast majority of errors (\emph{i.e.}, decays from the Rydberg state that is used to implement two-qubit gates) result in transitions out of the computational subspace into levels whose population can be continuously monitored using cycling transitions that do not disturb the qubit levels. As a result, the location of these errors is revealed, converting them into erasures. We estimate a fraction $R_e = 0.98$ of all errors can be detected this way. Second, we quantify the benefit of erasure conversion at the circuit level, using simulations of the surface code. We find that the predicted level of erasure conversion results in a significantly higher threshold, $p_{th} = 4.15\%$, compared to the case of pure depolarizing errors ($p_{th} = 0.937\%$). Finally, we find a faster reduction in the logical error rate immediately below the threshold. 

These results are directly relevant to near-term experiments, which have already demonstrated cooling and trapping of alkaline earth-like atoms such as Sr and Yb \cite{Norcia2018,Cooper2018,Saskin2019,barnes2021}, and in particular, $^{171}$Yb \cite{ma2021,jenkins2021}. The state-of-the-art entangling gate fidelity in neutral atoms \cite{Levine2019,madjarov2020} is already below the projected surface code threshold with erasure conversion. Furthermore, demonstrated scaling to hundreds of atoms should allow the encoding of many logical qubits at moderate code distances \cite{ebadi2021,scholl2021}.

Lastly, we note that our approach is complementary to a recent proposal for fault-tolerant computing with Rydberg arrays from Cong et.~al.~\cite{cong2021}, which is based on realizing highly biased noise and correcting leakage errors with additional ancilla operations. In comparison, the erasure conversion protocol in this work also handles leakage errors, but without requiring additional gates. Additionally, the circuit-level threshold for erasure errors is similar to or higher than that for biased noise, but does not restrict the circuit to bias-preserving gates.

\section{Erasure conversion in ${}^{171}$Yb qubits}
In a neutral atom quantum computer, an array of atomic qubits are trapped, manipulated and detected using light projected through a microscope objective (Fig.~\ref{fig:fig1}a). A variety of atomic species have been explored, but in this work, we consider ${}^{171}$Yb \cite{ma2021,jenkins2021}, with the qubit encoded in the $F=1/2$ $6s6p$ ${}^{3}$P$_{0}$ (Fig.~\ref{fig:fig1}b) level. This is commonly used as the upper level of optical atomic clocks \cite{ludlow2015}, and is metastable with a lifetime of $\tau\approx20$ s. We define the qubit states as $\ket{1} \equiv \ket{m_F=1/2}$ and $\ket{0} \equiv \ket{m_F=-1/2}$. Protocols for state preparation, measurement and single qubit rotations are presented in the supplementary information \cite{SI}, and we note that encoding the qubit in a metastable state confers other advantages for these operations, some of which have previously been discussed in the context of trapped ions \cite{allcock2021}. To perform two-qubit gates, the state $\ket{1}$ is coupled to a Rydberg state $\ket{r}$ with Rabi frequency $\Omega$. For concreteness, we consider the $6s75s$ $^3$S$_1$ state with $\ket{F,m_F} = \ket{3/2,3/2}$ \cite{Wilson2019}. Selective coupling of $\ket{1}$ to $\ket{r}$ can be achieved by using a circularly polarized laser and a large magnetic field to detune the transition from $\ket{0}$ to the $m_F=1/2$ Rydberg state \cite{ma2021}.

The resulting three level system $\{\ket{0},\ket{1},\ket{r}\}$ is analogous to hyperfine qubits encoded in alkali atoms, for which numerous gate protocols have been proposed and demonstrated \cite{jaksch2000,Lukin2001,isenhower2010,wilk2010,Levine2019,mitra2020,saffman2020}. These gates are based on the Rydberg blockade: the van der Waals interaction $V_{rr}(x) = C_6/x^6$ between a pair of Rydberg atoms separated by $x$ prevents their simultaneous excitation to $\ket{r}$ if $V_{rr}(x) \gg \Omega$. The gate duration is of order $t_g \approx 2\pi/\Omega \gg 2\pi/V_{rr}$, and during this time, the Rydberg state can decay with probability $p = \langle P_r \rangle \Gamma t_g$, where $\langle P_r \rangle \approx 1/2$ is the average population in $\ket{r}$ during the gate, and $\Gamma$ is the total decay rate from $\ket{r}$. This is the fundamental limitation to the fidelity of Rydberg gates \cite{Saffman2010}. It can be suppressed by increasing $\Omega$ (up to the limit imposed by $V_{rr}$), but in practice, $\Omega$ is often constrained by the available laser power.

The state $\ket{r}$ can decay via radiative decay to low-lying states (RD), or via blackbody-induced transitions to nearby Rydberg states (BBR) \cite{Saffman2010}. Crucially, a large fraction of RD events do not reach the qubit subspace $Q$, but instead go to the true atomic ground state $6s^2$ ${}^{1}$S$_{0}$ (with suitable repumping of the other metastable state, $6s6p$ ${}^{3}P_{2}$). For an $n=75$ $^3S_1$ Rydberg state, we estimate that 61\% of decays are BBR, 34\% are RD to the ground state, and only 5\% are RD to the qubit subspace. Therefore, a total of 95\% of all decays leave the qubit in disjoint subspaces, whose population can be detected efficiently, converting these errors into erasures. The remaining 5\% can only cause Pauli errors in the computational space---there is no possibility for leakage, as the $Q$ subspace has only two sublevels.

Decays to states outside of $Q$ can be be detected using fluorescence on closed cycling transitions that do not disturb atoms in $Q$. Population in the ${}^{1}$S$_{0}$ level can be efficiently detected using fluorescence on the ${}^{1}$P$_{1}$ transition at 399 nm \cite{yamamoto2016,Saskin2019} (subspace $R$ in Fig.~\ref{fig:fig1}c). This transition is highly cyclic, with a branching ratio of $\approx 1 \times 10^{-7}$ back into $Q$ \cite{loftus2000}. Population remaining in Rydberg states at the end of a gate can be converted into Yb$^+$ ions by autoionization on the $6s \rightarrow 6p_{1/2}$ Yb$^+$ transition at 369 nm \cite{burgers2021}. The resulting slow-moving Yb$^{+}$ ions can be detected using fluorescence on the same Yb$^{+}$ transition, as has been previously demonstrated for Sr$^+$ ions in ultracold strontium gases \cite{mcquillen2013} (subspace $B$ in Fig.~\ref{fig:fig1}c). As the ions can be removed after each erasure detection round with a small electric field, this approach also eliminates correlated errors from leakage to long-lived Rydberg states \cite{goldschmidt2016}. We estimate that site-resolved detection of atoms in ${}^{1}$S$_{0}$ with a fidelity $F > 0.999$ \cite{bergschneider2018}, and Yb$^+$ ions with a fidelity $F > 0.99$, can be achieved in a 10 $\mu$s imaging period \cite{SI}. We note that two nearby ions created in the same cycle will likely not be detected because of mutual repulsion, but this occurs with a very small probability relative to other errors, as discussed below.

\begin{figure}
    \centering
    \includegraphics{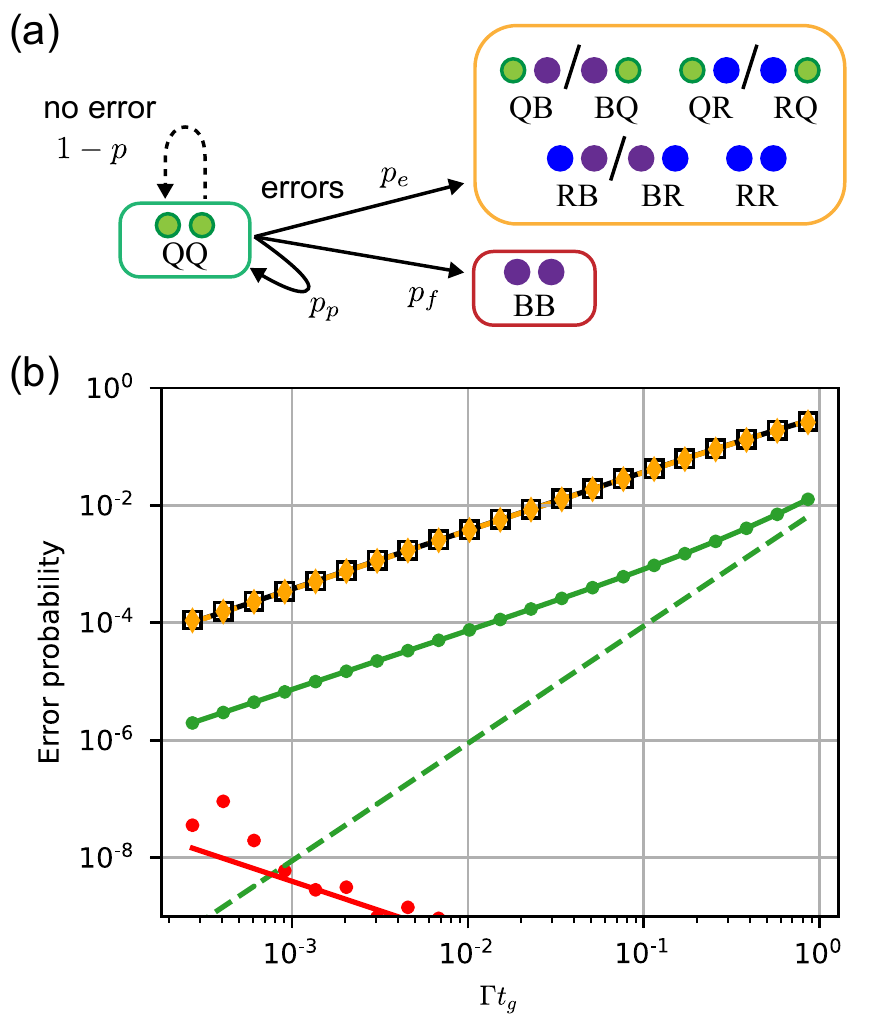}
    \caption{\label{fig:gatesim}(a) Possible atomic states at the end of a two-qubit gate. The configurations grouped in the yellow box are detectable erasure errors; red, undetectable errors; and green, the computational space. (b) Gate error as a function of the gate duration $t_g$. The average gate infidelity $1-\mathcal{F}$ (black squares) is dominated by detectable erasures with probability $p_e$ (orange points). The infidelity conditioned on not detecting an erasure, $1-\mathcal{F}_{\bar{e}}$ (green points) is about 50 times smaller. This reflects decays to $Q$ with probability $p_p$, and a no-jump evolution contribution (green dashed line). The probability $p_f$ of undetectable leakage (red points) is very small. The lines are analytic estimates of each quantity, while the symbols are numerical simulations. Both assume $V_{rr}/\Gamma = 10^6$, and $\Omega$ is varied along the horizontal axis \cite{SI}.
    }
\end{figure}

We divide the total spontaneous emission probability, $p$, into three classes depending on the final state of the atoms (Fig.~\ref{fig:gatesim}a). The first outcome is states corresponding to detectable erasures (BQ/QB, RQ/QR, RB/BR, and RR), with probability $p_e$. The second is the creation of two ions (BB), which cannot be detected, occurring with probability $p_f$. The third outcome is a return to the qubit subspace (QQ), with probability $p_p$, which results in a Pauli error on the qubits.

The value of $p$ and its decomposition depends on the specific Rydberg gate protocol. We study a particular example, the symmetric CZ gate from Ref. \cite{Levine2019}, using a combination of analytic and numerical techniques, detailed in the supplementary information \cite{SI} and summarized in Fig.~\ref{fig:gatesim}b. The probability of a detectable erasure, $p_e$, is almost identical to the average gate infidelity $1-\mathcal{F}$, indicating that the vast majority of errors are of this type. We infer the rate of Pauli errors on the qubits from the fidelity conditioned on \emph{not} detecting an erasure, $\mathcal{F}_{\bar{e}}$, as $p_p = 1-\mathcal{F}_{\bar{e}}$, and find $p_p \approx p_e/50$. Non-detectable leakage ($BB$) is strongly suppressed by the Rydberg blockade, and we find $p_f < 10^{-4} \times p_e$ over the relevant parameter range. Since decays occur preferentially from $\ket{1}$, continuously monitoring for erasures introduces an additional probability of gate error from non-Hermitian no-jump evolution \cite{plenio1998}, proportional to $p_e^2$, which is insignificant for $p_e < 0.1$.

We conclude that this approach effectively converts a fraction $R_e = p_e/(p_e+p_p) = 0.98$ of all spontaneous decay errors into erasures. This is a larger fraction than would be na\"ively predicted from the branching ratio into the qubit subspace, $1-\Gamma_Q/\Gamma = 0.95$, because decays to $Q$ in the middle of the gate result in re-excitation to $\ket{r}$ with a high probability, triggering an erasure detection. This value is in agreement with an analytic estimate \cite{SI}.

\section{Surface code simulations}

We now study the performance of an error correcting code with erasure conversion using circuit-level simulations. We consider the planar XZZX surface code \cite{bonillaataides2021}, which has been studied in the context of biased noise, and performs identically to the standard surface code for the case of unbiased noise. We implement Monte Carlo simulations of errors in a $d \times d$ array of data qubits to implement a code with distance $d$, and estimate the logical failure rate after $d$ rounds of measurements.

\begin{figure}
    \centering
    \includegraphics[width=3.5in]{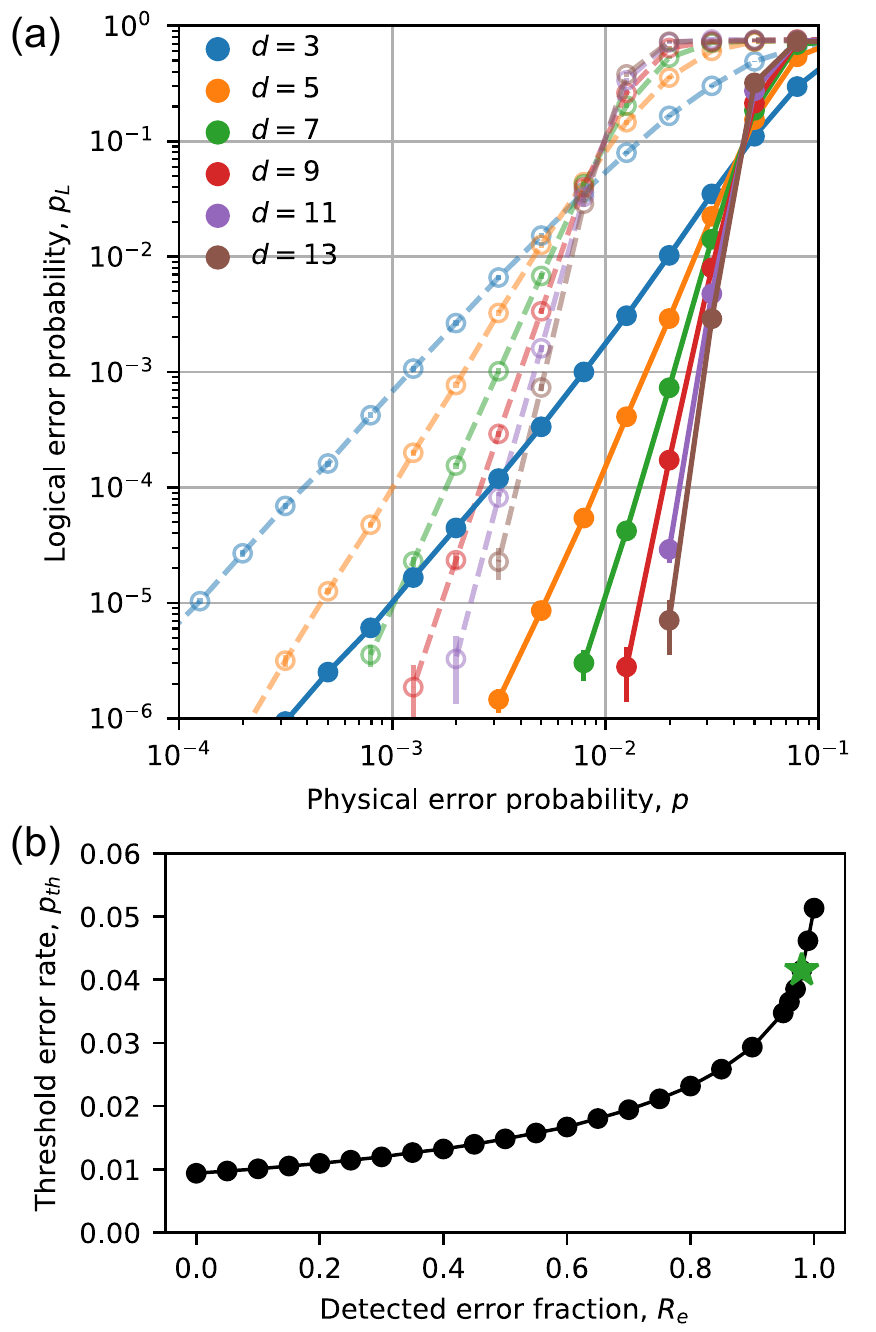}
    \caption{Circuit-level error thresholds in the presence of erasure errors. (a) Scaling of the logical error rate with the physical qubit error rate $p$ in the case of pure computational errors ($R_e=0$, open circles, dashed lines) and in the case of a high conversion to erasure errors, $R_e=0.98$ (filled circles, solid lines). The error thresholds are $p_{th} = 0.937(4)\%$ and $p_{th}=4.15(2)\%$, respectively, determined from the crossing of $d=11$ and $d=15$. The error bars indicate the 95\% confidence interval in $p_L$, estimated from the number of trials in the Monte Carlo simulation. (b) $p_{th}$ as a function of $R_e$ (The green star highlights $R_e=0.98$.)}
    \label{fig:pl}
\end{figure}

In the simulation, each two-qubit gate experiences either a Pauli error with probability $p_p = p(1-R_e)$, or an erasure with probability $p_e = p R_e$. The Pauli errors are drawn uniformly at random from the set $\{I,X,Y,Z\}^{\otimes 2}\backslash \{I\otimes I\}$, each with probability $p_p/15$. Following a two-qubit gate in which an erasure error occurs, both atoms are replaced with fresh ancilla atoms in a mixed state $I/2$ (Fig.~\ref{fig:fig1}d). We model this in the simulations by applying a Pauli error chosen uniformly at random from $\{I,X,Y,Z\}^{\otimes 2}$. We do not consider single-qubit gate errors or ancilla initialization or measurement errors at this stage \cite{SI}.

The syndrome measurement results, together with the locations of the erasure errors, are decoded with weighted Union Find (UF) decoder \cite{delfosse2021,huang2020a} to determine whether the error is correctable or leads to a logical failure. The UF decoder is optimal for pure erasure errors \cite{delfosse2020linear}, and performs comparably to conventional matching decoders for Pauli errors, but is considerably faster \cite{delfosse2021,huang2020a}.

In Fig.~\ref{fig:pl}a, we present the simulation results for $R_e=0$ and $R_e=0.98$. The former corresponds to pure Pauli errors, while the latter corresponds to the level of erasure conversion anticipated in $^{171}$Yb. The logical errors are significantly reduced in the latter case. The fault-tolerance threshold, defined as the physical error rate where the logical error rate decreases with increasing $d$, increases by a factor of 4.4, from $p_{th}=0.937\%$ to $p_{th} = 4.15\%$. In Fig.~\ref{fig:pl}b, we plot the threshold as a function of $R_e$. It reaches $5.13\%$ when $R_e=1$. The smooth increase of the threshold with $R_e$ is qualitatively consistent with previous studies of the surface code performance with mixed erasures and Pauli errors \cite{stace2009,barrett2010,delfosse2021}.

In addition to increasing the threshold, the high fraction of erasure errors also results in a faster decrease in the logical error rate below the threshold. Below the threshold, $p_L$ can be approximated by $A p^\nu$, where the exponent $\nu$ is the number of errors needed to cause a logical failure. A larger value of $\nu$ results in a faster suppression of logical errors below the threshold, and better code performance for a fixed number of qubits (\emph{i.e.}, fixed $d$).

\begin{figure}
    \centering
    \includegraphics[width=3.5in]{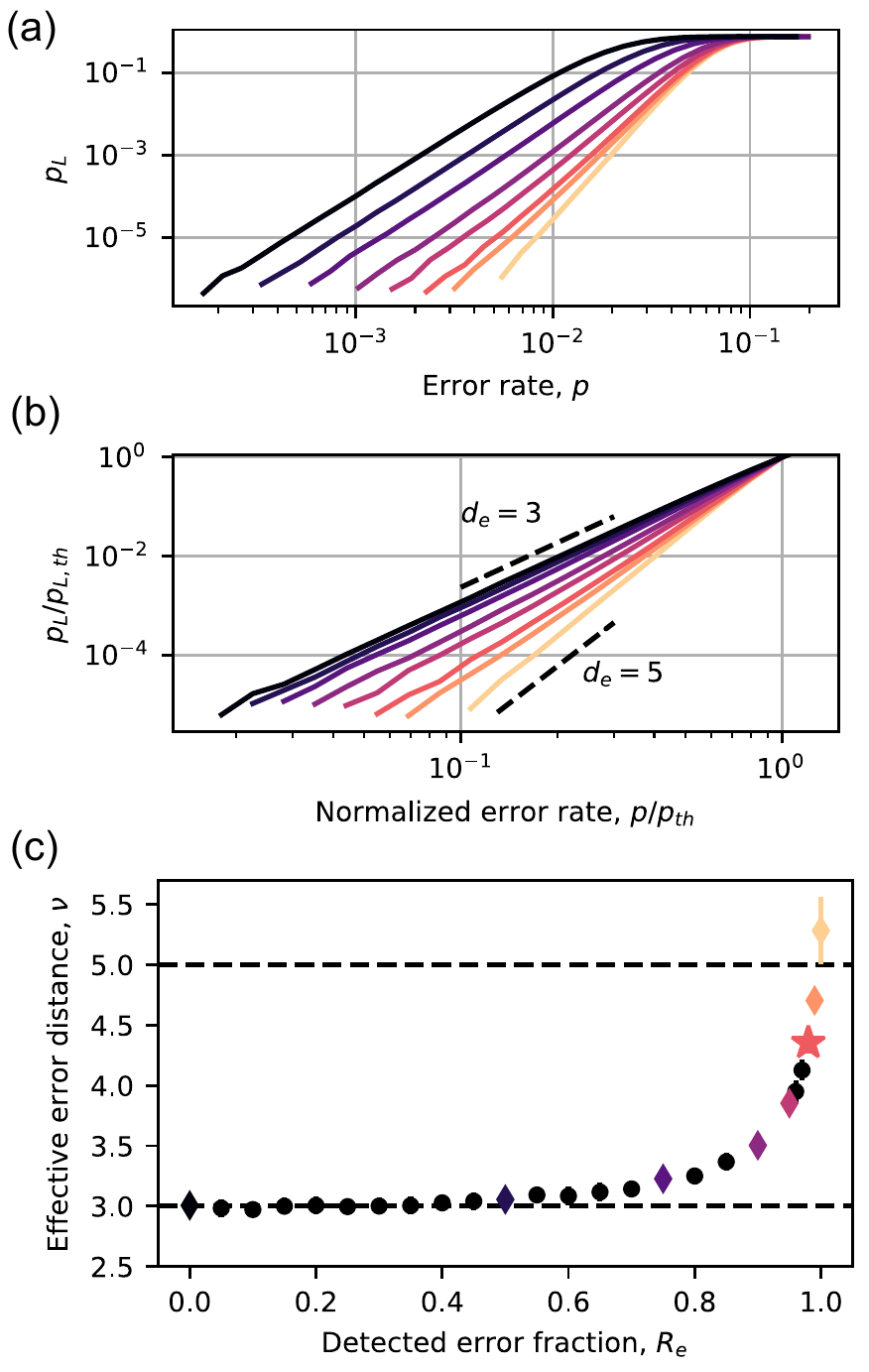}
    \caption{Logical error scaling below threshold. (a) $p_L$ vs $p$ at a fixed code distance $d=5$ for various values of $R_e$ [colors correspond to the diamond points in panel (c)]. In panel (b), the physical and logical error rates are rescaled by their values at the threshold. (c) Logical error exponent $\nu$, extracted from the slope of the curves in (b). The dashed lines show the expected asymptotic exponents for pure computational errors ($\nu_p=3$) and pure erasure errors ($\nu_e=5$).
}
    \label{fig:dist}
\end{figure}

In Fig.~\ref{fig:dist}a, we plot the logical error rate as a function of the physical error rate for a $d=5$ code for several values of $R_e$. When normalized by the threshold error rates (Fig.~\ref{fig:dist}b), it is evident that the exponent (slope) $\nu$ increases with $R_e$. The fitted exponents (Fig.~4c) smoothly increase from the expected value for pure Pauli errors, $\nu_p = (d+1)/2 = 3$, to the expected value for pure erasure errors, $\nu_e = d = 5$ (in fact, it exceeds this value slightly in the region sampled, which is close to the threshold). For $R_e=0.98$, $\nu = 4.35(2)$. Achieving this exponent with pure Pauli errors would require $d=7$, using nearly twice as many qubits as the $d=5$ code in Fig.~\ref{fig:dist}. For very small $p$, the exponent will eventually return to $\nu_p$, as the lowest weight failure ($\nu_p$ Pauli errors) will become dominant. The onset of this behavior is barely visible for $d=5$ in Fig.~\ref{fig:pl}a.

\section{Discussion}

There are several points worth discussing. First, we note that the threshold error rate for $R_e=0.98$ corresponds to a two-qubit gate fidelity of 95.9\%, which is exceeded by the current state-of-the-art. Recently, entangled states with fidelity $\mathcal{F} = 97.4\%$ were demonstrated for hyperfine qubits in Rb \cite{Levine2019}, and we also note that $\mathcal{F} = 99.1\%$ has been demonstrated for ground-Rydberg qubits in ${}^{88}$Sr \cite{madjarov2020}. With reasonable technical improvements, a reduction of the error rate by at least one order of magnitude has been projected \cite{saffman2020}, which would place neutral atom qubits far below the threshold, into a regime of genuine fault-tolerant operation. Arrays of hundreds of neutral atom qubits have been demonstrated \cite{ebadi2021,scholl2021}, which is a sufficient number to realize a single surface code logical qubit with $d=11$, or five logical qubits with $d=5$. While we analyze the surface code in this work because of the availability of simple, accurate decoders, we expect erasure conversion to realize a similar benefit on any code. In combination with the flexible connectivity of neutral atom arrays enabled by dynamic rearrangement \cite{beugnon2007,yang2016,bluvstein2021}, this opens the door to implementing a wide range of efficient codes \cite{breuckmann2021}.

Second, in order to compare erasure conversion to previous proposals for achieving fault-tolerant Rydberg gates by repumping leaked Rydberg population in a bias-preserving manner \cite{cong2021}, we have also simulated the XZZX surface code with biased noise and bias-preserving gates. For noise with bias $\eta$ (\emph{i.e.}, if the probability of $X$ or $Y$ errors is $\eta$ times smaller than $Z$ errors), we find a threshold of $p_{th}=2.27\%$ for the XZZX surface code when $\eta = 100$, which increases to $p_{th} = 3.69\%$ when $\eta\rightarrow \infty$. For comparison, the threshold with erasure conversion is higher than the case of infinite bias if $R_e \geq 0.96$.

Third, our analysis has focused on two-qubit gate errors, since they are dominant in neutral atom arrays, and are also the most problematic for fault-tolerant error correction \cite{fowler2012a}. However, with very efficient erasure conversion for two-qubit gate errors, the effect of single-qubit errors, initialization and measurement errors, and atom loss may become more significant. In the supplementary information, we present additional simulations showing that the inclusion of initialization, measurement, and single-qubit gate errors with reasonable values does not significantly affect the threshold two-qubit gate error. We also note that erasure conversion can also be effective for other types of spontaneous errors, including Raman scattering during single qubit gates, the finite lifetime of the $^3$P$_0$ level, and certain measurement errors. Atom loss can occur spontaneously (\emph{i.e.}, from collision with background gas atoms) or as a result of an undetected erasure, but these probabilities are both very small compared to $p$. In this regime, these undetected leakage events can be handled fault-tolerantly with only one extra gate per stabilizer measurement, with very small impact on $p_{th}$ \cite{suchara2015}. We leave a detailed analysis to future work.

Lastly, we highlight that erasure conversion can lead to more resource-efficient, fault-tolerant subroutines for universal computation, such as magic-state distillation~\cite{bravyi2005universal}. This protocol uses several copies of faulty resource states to produce fewer copies with lower error rate. This is expected to consume large portions of the quantum hardware \cite{fowler2012a,fowler2013surface}, but the overhead can be reduced by improving the fidelity of the input raw magic states. By rejecting resource states with detected erasures, the error rate can be reduced from $O(p)$ \cite{horsman2012surface,landahl2014quantum,li2015magic,luo2021quantum} to $O((1-R_e)p)$. Therefore, $98\%$ erasure conversion can give over an order of magnitude reduction in the infidelity of raw magic states, resulting in a large reduction in overheads for magic state distillation.

\section{Conclusion}

We have proposed an approach for efficiently implementing fault-tolerant quantum logic operations in neutral atom arrays using ${}^{171}$Yb. By leveraging the unique level structure of this alkaline earth atom, we convert the dominant source of error for two-qubit gates---spontaneous decay from the Rydberg state---into directly detected erasure errors. We find a 4.4-fold increase in the circuit-level threshold for a surface code, bringing the threshold within the range of current experimental gate fidelities in neutral atom arrays. Combined with a steeper scaling of the logical error rate below the threshold, this approach is promising for demonstrating fault-tolerant logical operations with near-term experimental hardware. We anticipate that erasure conversion will also be applicable to other codes and other physical qubit platforms.

\section{Acknowledgements}
We gratefully acknowledge Alex Burgers, Shuo Ma, Genyue Liu, Jack Wilson, Sam Saskin and Bichen Zhang for helpful conversations, and Ken Brown, Steven Girvin and Mark Saffman for a critical reading of the manuscript. SP and JDT acknowledge support from the National Science Foundation (QLCI grant OMA-2120757). JDT also acknowledges additional support from ARO PECASE (W911NF-18-10215), ONR (N00014-20-1-2426), DARPA ONISQ (W911NF-20-10021) and the Sloan Foundation. SK acknowledges support from the National Science Foundation (QLCI grant OMA-2016136) and the ARO (W911NF-21-1-0012).

\bibliography{qec.bib,addon.bib}

\section{Methods}
\subsection{Error correcting code simulations}

In this section, we provide additional details about the simulations used to generate the results shown in Figures \ref{fig:pl} and \ref{fig:dist}. We assign each two-qubit gate to have an error from the set $\{I,X,Y,Z\}^{\otimes 2}\backslash \{I\otimes I\}$ with probability $p_p/15$, and an erasure error with probability $p_e$, with $p_e/(p_p+p_e) = R_e$. Immediately after an erasure error on a two-qubit gate, both qubits are re-initialized in a completely mixed state which is modelled using an error channel $(I\rho I+X\rho X+Y\rho Y+Z\rho Z)/4$ on each qubit. In the case of an erasure, the qubit is replaced with a completely mixed state and the recorded measurement outcome is random. We choose this model for simplicity, but in the experiment, better performance may be realized using an ancilla polarized into $\ket{1}$, as Rydberg decays only happen from this initial state. We assume the existence of native CZ and CNOT gates, so a stabilizer cycle can be completed without single-qubit gates. We also neglect idle errors, since these are typically insignificant for atomic qubits.

Ancilla initialization (measurement) are handled in a similar way, with a Pauli error following (preceding) a perfect operation, with probability $p_m$ ($p_m=0$ in Figs.~\ref{fig:pl}, \ref{fig:dist}, but results for $p_m>0$ are shown in Figure \ref{fig:pm_si}). We assume the existence of native CZ and CNOT gates, so a stabilizer cycle can be completed without single-qubit gates. We also neglect idle errors, since these are typically insignificant for atomic qubits.

We simulate the surface code with open boundary conditions. Each syndrome extraction round proceeds in six
steps: ancilla state preparation, four two-qubit gates applied in the order shown in Fig.~\ref{fig:fig1}, and finally a
measurement step. For a $d\times d$ lattice, we perform $d$ rounds of syndrome measurements, followed by one final round of perfect measurements. The decoder
graph is constructed by connecting all space-time points generated by errors in the circuit applied as discussed above. Each
of these edges is then weighted by $\mathrm{ln}(p')$ truncated to the nearest integer, where $p'$ is the largest single error probability that gives rise to the edge. After sampling an error, the weighted UF decoder is applied
to determine error patterns consistent with the syndromes. We do not apply the peeling decoder but account for logical errors by keeping track of parity of defects crossing the logical boundaries. Our implementation of the decoder was separately benchmarked against the results in~\cite{huang2020a} and yields same thresholds.

For the comparison to the threshold of the XZZX code when the noise is biased, we apply errors from $Q= \{I,X,Y,Z\}^{\otimes 2}\backslash \{I\otimes I\}$ after two qubit gate with probability $p_Q$. The first (second) operator in the tensor product is applied to the control (target) qubit. We assume bias-preserving CNOT gates and thus use $p_{ZI}=p$, $p_{IZ}=p_{ZZ}=p/2$ with the probability of other non-pure-dephasing Pauli errors $=p/\eta$ \cite{darmawan2021}. For the CZ gate we use $p_{ZI}=p$, $p_{IZ}=p$ with the probability of other non-pure-dephasing Pauli errors $=p/\eta$. For the threshold quoted in the main text no single-qubit preparation and measurement noise is applied, to facilitate direct comparison to the threshold with erasure conversion in Fig.~\ref{fig:pl}. In the main text we quote threshold in terms of the total two-qubit gate infidelity $\sim 2p$ for large $\eta$, to facilitate comparison to the threshold in Fig.~\ref{fig:pl}.

\clearpage

\setcounter{section}{0}
\renewcommand{\thesection}{S\arabic{section}}
\renewcommand\theequation{S\arabic{equation}}
\renewcommand\thefigure{S\arabic{figure}}
\renewcommand\thetable{S\arabic{table}}

\begin{center}
    {\LARGE Supplementary Information}
\end{center}

\section{$^{171}$Yb gate operations}
\label{app:protocols}
Here, we provide a sketch of a universal set of gate operations on qubits encoded in the ${}^3$P${}_0$ level of $^{171}$Yb (recently, universal operations were demonstrated on ground state $^{171}$Yb qubits \cite{ma2021}). Starting with an atom in $^1$S$_0$, initialization into $\ket{1}$ can be performed by optically pumping into $\ket{^1\text{S}_0, m_F=1/2}$ and transferring to the $^3$P$_0$ (manifold Q) using the clock transition. Mid-circuit measurement can be performed using the same clock pulse to selectively transfer population in $\ket{1}$ to $^1$S$_0$, and measuring the ${}^1$S${}_0$ population with fluorescence. As an alternative to driving the clock transition, optical pumping via intermediate $S$ and $D$ states can also be used.

Single qubit gate rotations can be performed using Raman transitions and light shifts on the $6s7s$ $^3$S$_1$ transition (649 nm), or via the Rydberg state. In both cases, errors can arise from photon scattering, but erasure conversion can be performed at a similar or greater level than for the two-qubit gates discussed in the main text (see section \ref{app:othererrors}).

\section{Yb Branching Ratios}
In this section, we consider the decay pathways from the Rydberg state, which determine the probability that a spontaneous decay is converted into an erasure. These calculations involve dipole matrix elements between ground states and Rydberg states in Yb that have not been directly measured or computed with rigorous many-body techniques. Therefore, we estimate them using a single active electron approximation \cite{bates1949}, and wavefunctions computed using the Numerov technique \cite{weber2017}. We focus on the $6s75s$ $^3$S$_1$ $F=3/2$ state for concreteness \cite{Wilson2019}.

\begin{figure}
    \centering
    \includegraphics[width=3.5in]{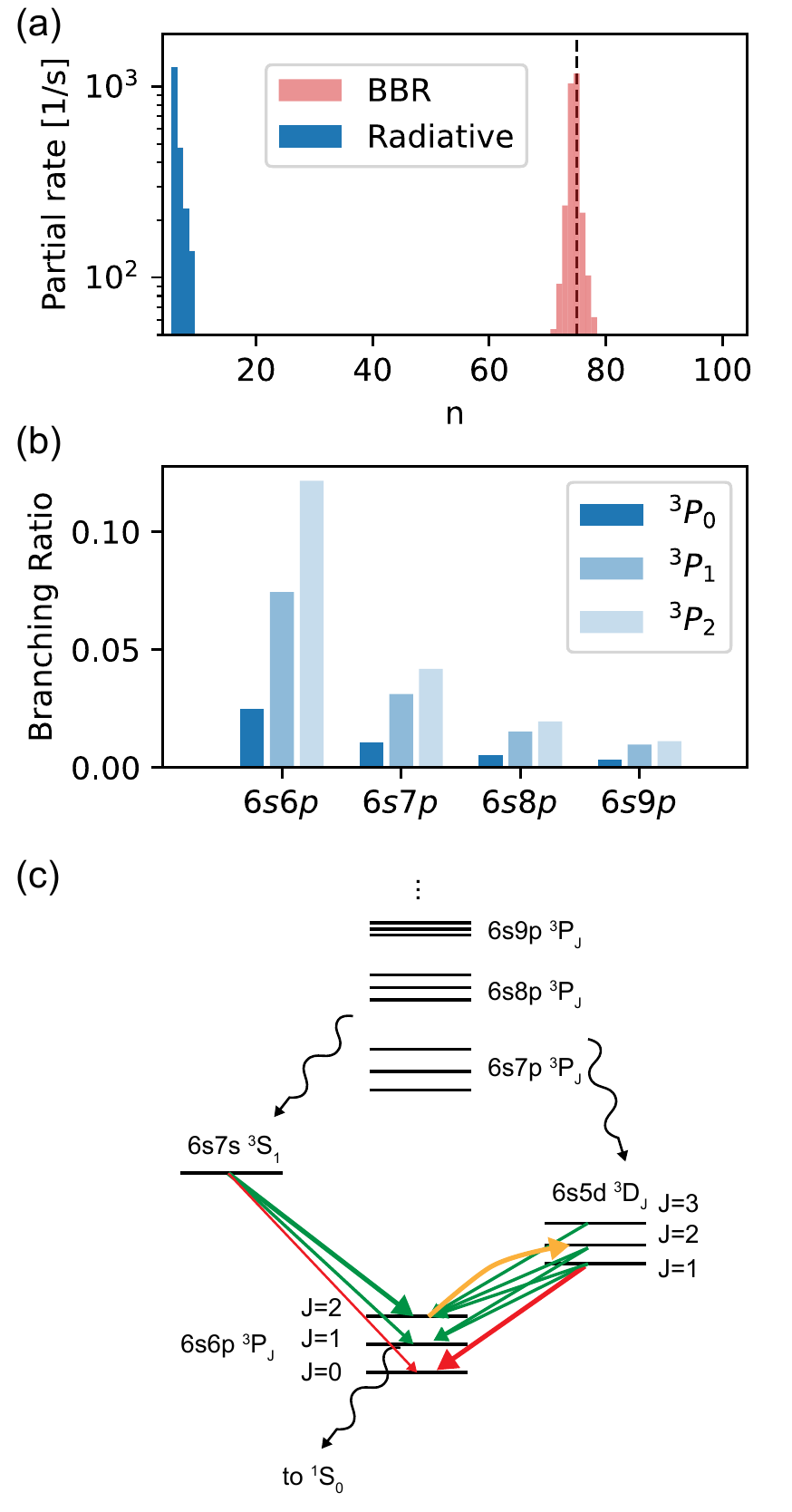}
    \caption{Decay pathways from the Yb $^3$S$_1$ Rydberg series. (a) Partial decay rates to all $P$ states of a given principal quantum number $n$, starting from $n=75$. BBR transitions to nearby states are shown in red, and radiative decay to lower $n$ states are shown in blue. Final states with $n=10-35$ are not included because of the absence of spectroscopic data. (b) Branching ratio into low-energy $6sps$ $^3$P$_J$ states. The branching ratio into the qubit manifold Q, $6s6p$ $^3$P$_0$, is 2.5\%. (c) Partial energy diagram showing relevant transitions between low-lying states. Decays in red are to $Q$, while the orange transition can be used to repump $^3P_2$ without populating $Q$.}
    \label{fig:decays}
\end{figure}

The decay pathways can be separated into BBR decays to nearby $n$ and radiative decays to low-$n$ states. For $n=75$, the BBR decay rate is 3480 1/s, and the radiative decay rate is 2200 1/s, which gives a branching ratio of 0.39 into radiative decay, and 0.61 into BBR decay (Fig.~\ref{fig:decays}a).

The radiative decays favor the lowest energy states, because of the larger density of states at the relevant transition energy \cite{Saffman2010}. However, angular momentum algebra favors higher $J$ states within the same fine structure manifold. Therefore, the fraction of decays that terminate directly in the $J=0$ qubit manifold Q is only 0.025 (Fig.~\ref{fig:decays}b).

Decay events to $6s6p$ $^3$P$_1$ will quickly relax to the ground state $^1$S$_0$ via a second spontaneous decay. Decays to $6s6p$ $^3$P$_2$ can be repumped to $6s6p$ $^3$P$_1$ via $6s5d$ $^3$D$_2$, which cannot decay to the qubit subspace because of angular momentum selection rules.

However, approximately 0.17 of all the decay events are to $6snp$ states with $n>6$. These states will overwhelmingly decay to the $6s7s$ $^3$S$_1$ and $6s5d$ $^3$D$_J$ states, which in turn can decay to $6s6p$ states (Fig.~\ref{fig:decays}c). No data is available to estimate the relative branching ratio between the $S$ and $D$ decay pathways, but we can estimate the fraction of decays that return to Q within each pathway.

The state $6s7s$ $^3$S$_1$ decays into the $6s6p$ $^3$P$_J$ levels with a branching ratio that can be estimated as \cite{martin2013}:
\begin{equation}
\label{eq:we}
    \frac{\Gamma_J}{\Gamma_{tot}} = \frac{1}{\mathcal{N}}\omega^3_J (2J+1)(2L'+1) \sixj{L}{L'}{1}{J'}{J}{S}^2
\end{equation}

Here, the primed quantities denote the angular momenta of the initial state ($^3$S$_1$), and the unprimed quantitites for the final state ($^3$P$_J$). $\omega_J$ is the transition frequency for the decay to the state $J$, and the normalization constant $\mathcal{N}$ ensures $\sum_J \Gamma_J = \Gamma_{tot}$. The branching ratios into $J = \{0,1,2\}$ are $\{0.15,0.40,0.45\}$. Therefore, around 0.15 of all decays via $^3$S$_1$ will reach Q.

In the case of decays via the $6s5d$ $^3$D$_J$ states, we can use Eq. \eqref{eq:we} to estimate the branching ratio from $6snp$ $^3$P$_J'$ to the various $^3$D$_J$ states. Since only $^3$D$_1$ can decay to the Q, we only state this fraction, which is approximately $\{1,0.25,0.01\}$ when starting from $6snp$ $^3P_{J'}$ with $J'=\{0,1,2\}$. To estimate the branching ratio from $6s5d$ $^3$D$_1$ to Q, we do not use Eq. \eqref{eq:we} because the states are rather close in energy, but instead use the theoretical matrix elements in Ref. \cite{porsev1999}, which give a branching ratio of 0.65. Combining this with the distribution of population among the $6snp$ $^3P_{J'}$ levels in Fig.~\ref{fig:decays}b, we arrive at an estimate that 0.16 of the decays via $D$ states terminate in Q.

As the probability to end up in Q via the $S$ or $D$ decay pathways is similar, the (unknown) branching ratio between them becomes unimportant. Taking it to be 0.5, we conclude that 14\% of decays from $6snp$ levels with $n>6$ return to $Q$. Adding this to the direct decays to $Q$, we arrive at a final estimate that 0.051 of all Rydberg decays return to the qubit manifold Q.

Lastly, we note that this analysis does not include the effect of doubly-excited states that perturb the Rydberg series, which can give rise to additional decay pathways \cite{vaillant2014a}. In Yb, these are especially prominent because of the number of core excited states \cite{aymar1984}. There is not enough spectroscopic data about the Yb Rydberg series to quantitatively evaluate the impact of series perturbers. However, we note that these doubly excited states will require a minimum of three spontaneous decays to reach the $6s6p$ $^3$P$_J$ states. Given the general propensity to decay to higher $J$ states at each step, it is likely that the branching ratio into $^3$P$_0$ from doubly-excited perturbers will not be worse than the values estimated above.

We also do not explicitly include hyperfine structure in these calculations, but rather calculate matrix elements between $J$ states in $^{174}$Yb. This is an excellent approximation for the transitions from low-$n$ to Rydberg states, since these matrix elements are mainly sensitive to the Rydberg state quantum defect, and the $^3$S$_1$ $F=3/2$ Rydberg state that we consider has the same quantum defect as the $^3$S$_1$ series in $^{174}$Yb \cite{ma2021} because its core electron configuration is purely Yb$^+$ $F=1$. However, it is possible that the BBR transition rate varies slightly between isotopes, since the hyperfine splitting changes the energy level spacing by a significant amount. We believe that the error from this approximation is much less than the uncertainty arising from unknown series perturbers.

\section{Erasure detection fidelity}
\label{app:detection}
\subsection{Detection of atoms in $^1S_0$}
We first consider the localized detection fidelity for atoms $^1S_0$, using the cycling transition in the R manifold. Many protocols for imaging atoms in tweezers focus on non-destructive detection, and therefore image slowly while simultaneously cooling, which is not optimal for minimizing computational cycle time \cite{Saskin2019}. Here we instead consider rapid but destructive detection \cite{bergschneider2018}, with the aim of replacing atoms from a reservoir when erasures are detected (which occurs with a low probability). To estimate the fidelity, we take the atoms to be initially at rest, ignore the dipole trap, and assume illumination by counter-propagating fields above saturation, such that the photon scattering rate is $\Gamma/2$. This results in no net force on the atom, but momentum diffusion from photon recoils leads to an increasing mean squared atomic displacement of \cite{joffe1993,bergschneider2018}:
\begin{equation}
\langle x^2(t) \rangle  = \frac{v_{rec}^2}{3} \frac{t^3}{3} \frac{\Gamma}{2} = \frac{\hbar^2 k^2}{18 m^2} t^3 \Gamma
\end{equation}
where $\Gamma = 2\pi \times 28$ MHz is the $^1S_0$ - $^1P_1$ transition linewidth, the wavevector $k=2\pi/\lambda$ with $\lambda=399$ nm, and $m$ is the atomic mass.

We envision a tweezer array with a spacing of $a=3-5\,\mu$m, and therefore require that $\sqrt{\langle x^2(t) \rangle} < a/2$ to determine which site is fluorescing. In Ref.~\cite{bergschneider2018}, free-space imaging of single ${}^{6}$Li atoms was demonstrated with a detection fidelity of 99.4\% after an imaging time of 20 \textmu s, after which time $\sqrt{\langle x^2(t) \rangle}^{1/2} = 10.4$~\textmu m. During this time, approximately 330 photons were scattered, and 25 detected, with an EMCCD and a modest numerical aperture objective (NA=0.55). However, for the same number of detected photons, the position spread scales as $1/(m \lambda \Gamma)$ \cite{bergschneider2018}, and this quantity is a factor of 81 smaller for the heavy ${}^{171}$Yb compared to ${}^{6}$Li, so we anticipate a position spread of only 120 nm for the same conditions. Therefore, achieving imaging fidelity greater than 99.9\% should be readily achievable for atoms in ${}^{1}S_0$, in less than half the time, since the scattering rate for Yb is more than 3 times larger.

\subsection{Detection of ions}
We now consider the detection fidelity of Yb$^{+}$ ions using the cycling transition in manifold B following autoiniozation out of a Rydberg state. Ions created from Rydberg atoms have been imaged using fluorescence in ultracold quantum gases of strontium \cite{mcquillen2013}. Compared to detecting neutral atoms, there are two additional factors to consider: an initial velocity $v_0$ arising from recoil momentum from the ejected electron, and acceleration due to a background electric field or the presence of other ions. We begin by considering the initial velocity: when a $6p_{1/2}np$ Rydberg state decays to Yb$^+$ $(6s)$ + $e^-$ via autoionization, the electron carries away an energy $\Delta E \approx I_{6p_{1/2}} - I_{6s} \approx 27100\, \textrm{cm}^{-1}$, where $I_{j}$ is the ionization limit for Yb$^0$ corresponding to the ion core in state $j$, and we have made the approximation that the electron mass is very small compared to the ion mass. In this case, the ion acquires a recoil momentum $p_{e} = \sqrt{2 \Delta E m_e}$, corresponding to a velocity $v_0 = p_{e}/m \approx 3.5$ m/s.

With a finite initial velocity, the mean squared position is:
\begin{equation}
\langle x^2(t) \rangle = v_0^2 t^2 + \frac{\hbar^2 k^2}{18 m^2} t^3 \Gamma = v_{rec}^2 t^2 \left[\left(\frac{v_0}{v_{rec}}\right)^2 + \frac{t \Gamma}{18}\right]
\end{equation}
where $v_{rec} = \hbar k/m$ is the recoil velocity for the imaging wavelength, now 369 nm, and $\Gamma = 2\pi \times 19$ MHz. For the parameters above, $v_0/v_{rec} \approx 550$. Recognizing that the number of scattered photons is $N_{ph} = t \Gamma/2$, it is clear that the first term dominates for $N_{ph} < 10^6$. Therefore, we can express the position as:
\begin{equation}
\sqrt{\langle x^2(t) \rangle} = \frac{2 v_0 N_{ph}}{\Gamma} \approx 54\, \textrm{nm/photon}
\end{equation}
With a total detection efficiency of $\eta = 0.1$, an average of 5 photons can be detected while maintaining  $\sqrt{\langle x^2(t) \rangle } < 2.5\,\mu$m, corresponding to 99\% detection fidelity in the absence of dark counts. The necessary imaging time is less than 2 $\mu$s.

Achieving this collection and detection efficiency is challenging but achievable, for example, with a detector with 30\% quantum efficiency and NA=0.7 objectives from two sides. However, this is a pessimistic estimate of the requirements for several reasons. First, Yb$^+$ ions are only produced on atoms undergoing a two-qubit gate, and these gates cannot be performed on every atom in the array in parallel because of cross-blockade effects. Therefore, it is only necessary to resolve the atoms participating in gates in a particular cycle, which may have a separation of $2a$ or $3a$, allowing for longer imaging times and more particle spread. Second, we have assumed that the recoil momentum is always in the plane of the array. However, it is actually distributed in three dimensinos, and out-of-plane motion does not matter on the relevant time scale. Lastly, we have treated all autoionization events as transitions to Yb$^+$ $(6s)$, while in reality, a significant fraction of autoionization events will decay to a Yb$^+$ $(5d)$ state. These states can be quickly repumped to 6s, so imaging can proceed as normal. However, for this decay process, $\Delta E$ is smaller by a factor of approximately 6, and $v_0$ is smaller by a factor of 2.6.

We can also consider the role of a background electric field, which will cause a position displacement:

\begin{equation}
\Delta x = \frac{q E}{2 m} t^2 = \frac{q E}{2 m} \left(\frac{2 N_{ph}}{\Gamma}\right)^2
\end{equation}

Here, $E$ is the field strength and $q$ is the electron charge. Using the Yb$^{+}$ ion parameters and $N_{ph} = 200$, this results in a drift of approximately 316 nm/(mV/cm) during the imaging time. With intra-vacuum electrodes, it is possible to null background electric fields at the level of approximately 1 mV/cm \cite{Wilson2019}, so this is not a significant source of imaging error.

Lastly, we consider electric fields resulting from the simultaneous creation of multiple Yb$^{+}$ ions in a single gate cycle. An ion at a distance of $d=30\,\mu$m produces an electric field of 16 mV/cm, which will cause a displacement on both ions of approximately 5 $\mu$m during the time it takes to scatter 200 photons. Therefore, ion creation in a smaller radius will likely accelerate the ions too much to resolve their positions. This may motivate further reduction of the number of gates applied per cycle in the array.

\subsection{Alternate detection strategies for population in Rydberg states}
Fluorescence detection of ions has the benefit of being fast and compatible with existing experimental techniques. One alternate approach is to detect Yb$^+$ ions and electrons using charged particle optics and detectors. A second alternative is to simply wait for any Rydberg atoms to decay. To ensure more than 99.9\% of the ions have decayed, it would be necessary to wait approximately $\tau > 7/\Gamma \approx 1$ ms, and avoiding atom loss during this time will require that all of the intermediate Rydberg states are trapped. However, this is straightforward in alkaline earth atoms using the polarizability of the ion core \cite{Wilson2019}. Because of the large number of intermediate Rydberg states and their complex radiative decay pathways, it is not possible to accurately calculate the ultimate branching ratio back into $^3$P$_0$, but a crude estimate suggests it would result in less efficient erasure conversion, with $R_e \approx 0.9$.

\section{Gate Simulations}
\label{app:gatesim}

\begin{figure}
    \centering
    \includegraphics[width=3.5in]{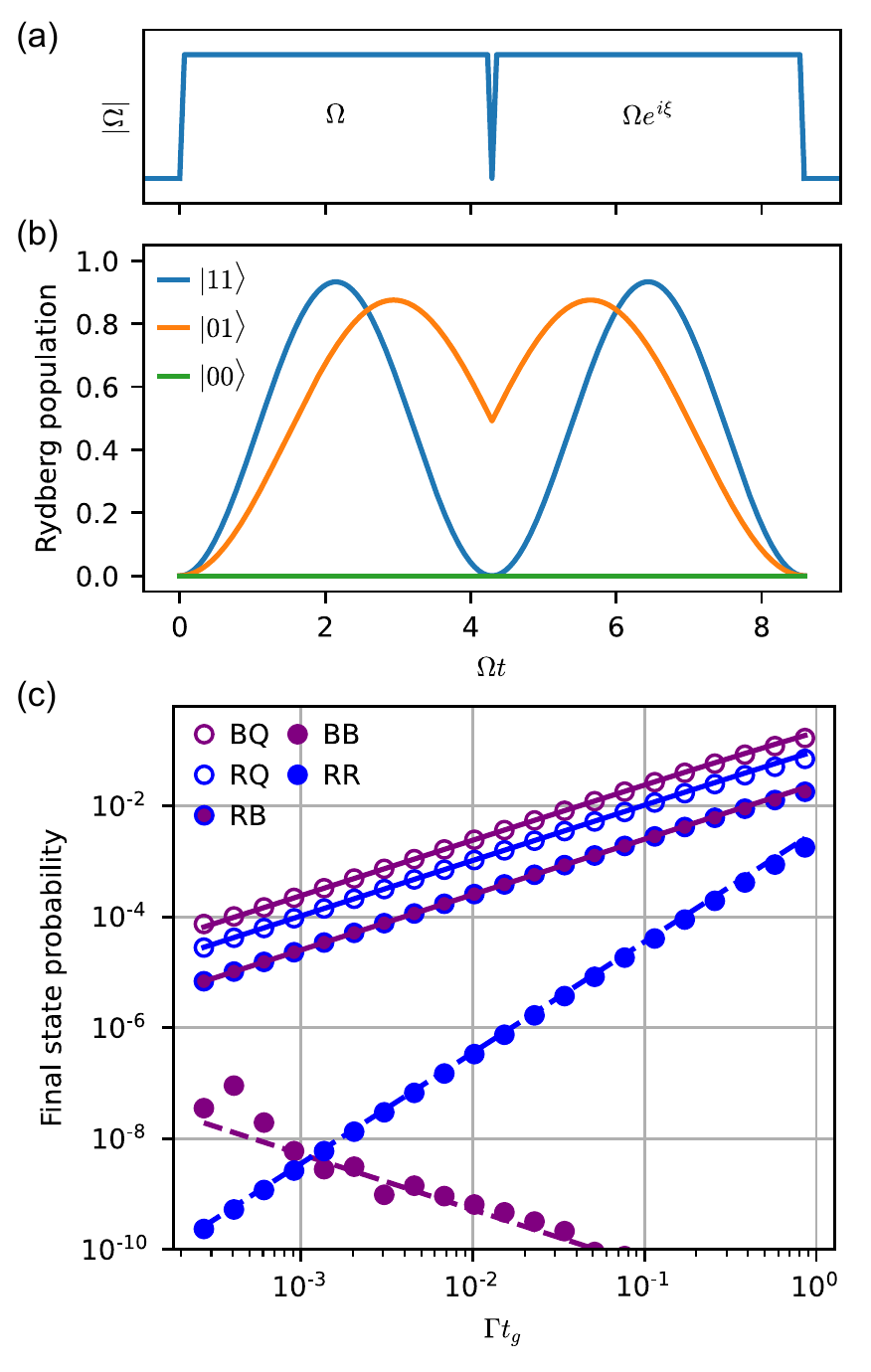}
    \caption{Gate simulations. (a) Pulse sequence used to implement the CZ gate from Ref. \cite{Levine2019}. (b) Rydberg state population during the gate, for various initial states. (c) Probability of individual erasure error channels (see Fig.~2a). The solid lines are analytic estimates from section \ref{app:gatesim}.}
    \label{fig:gatesim_si}
\end{figure}

In this section, we describe a detailed, microscopic simulation of a two-qubit gate using the level structure in Fig.~1, to evaluate the quantitative performance of the erasure conversion approach. While we expect that this protocol should work for any Rydberg gate, we focus specifically on the protocol introduced in Ref. \cite{Levine2019}, and applied to $^{171}$Yb in Ref. \cite{ma2021}, which we refer to hereafter as the \emph{LP gate}.

The system is described by the following two-atom Hamiltonian:

\begin{equation}
\begin{split}
    H &= \sum_{i=\{1,2\}} \frac{1}{2} \left(\Omega \ket{r}_{ii} \bra{1} + \Omega^* \ket{1}_{ii} {\bra{r}}\right) + \Delta \ket{r}_{ii}\bra{r} \\
    &+ V_{rr} \ket{rr}\bra{rr} + V_{pp} \ket{pp}\bra{pp} + V_{rp} (\ket{rp}\bra{pr} + h.c.)
\end{split}
\label{eq:ham}
\end{equation}

The qubit state $\ket{1}$ in each atom is coupled to $\ket{r}$ by a drive $\Omega$ with detuning $\Delta$. The Rydberg blockade shifts the state $\ket{rr}$ by $V_{rr}$. We also incorporate a single additional state, $\ket{p}$, that is populated by BBR transitions. This state has a a self-blockade interaction with strength $V_{pp}$, and a cross-blockade interaction with $\ket{r}$ with strength $V_{rp}$. Only states with large matrix elements to $\ket{r}$ are populated by BBR transitions, and therefore, $V_{rp}$ is dominated by the strong dipole-dipole interaction. Therefore, we expect that $V_{rp} \gg V_{pp},V_{rr}$.

The LP gate protocol is based on the fact that, when $V_{rr} \gg \Omega$, the initial state $\ket{11}$ cannot be excited to $\ket{rr}$, but is instead excited to $\ket{W} = (\ket{1r} + \ket{r1})/\sqrt{2}$ at a rate $\sqrt{2}\Omega$. Therefore, the use of an appropriate detuned pulse with a phase slip allows for excitation trajectories for all initial states that return to themselves, but with different accumulated phases for $\ket{11}$ and $\ket{01}$ (or $\ket{10}$), giving rise to a controlled-$Z$ (CZ) gate (Fig.~\ref{fig:gatesim_si}a,b) \cite{Levine2019}.

As discussed in the main text, the dominant, fundamental source of error is decay from $\ket{r}$ during the gate. This can result in a BBR transition to another Rydberg state, a radiative decay to the ground state $^1$S$_0$ ($\ket{g}$) or the computational level. During the erasure detection step, these correspond to three distinct outcomes: ion fluorescence (which we abbreviate $B$), ground state fluorescence ($R$) or no signal, indicating that the qubit remains in the computational space ($Q$). In a two-qubit gate, the outcome $QQ$ signals no erasure, while any other outcome is considered to be an erasure error on both qubits.

\subsection{Analytic error model}
\label{app:analyticerrors}

In this section, we derive analytic expressions for the probabilities of various errors to occur during the two-qubit gate. For atoms beginning in the state $\ket{00}$, there is no excitation to the Rydberg state, and therefore no errors. Below, we consider the other initial states.

\subsubsection{Initial state $\ket{01}$ (or $\ket{10}$)}

First, consider the case that the atoms start in $\ket{01}$. The case $\ket{10}$ is identical because the gate is symmetric in the two atoms. During the gate, in the absence of errors, we can represent the state of the atoms as:
\begin{equation}
    \ket{\psi(t)} = \psi_{1}(t)\ket{01} + \psi_r(t) \ket{0r}
\end{equation}
The Rydberg excitation probability $|\psi_r(t)|^2$ is plotted in Fig.~\ref{fig:gatesim_si}b.

\begin{figure}
    \centering
    \includegraphics{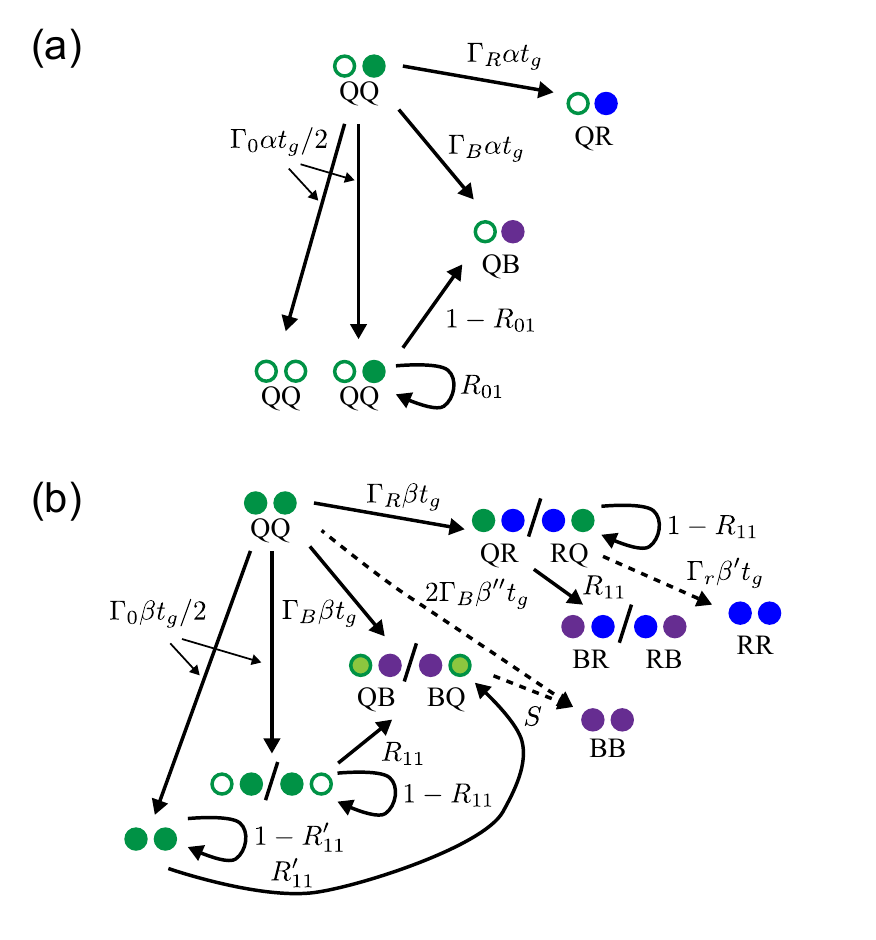}
    \caption{Diagram of transition probabilities during a two-qubit gate, for atoms beginning in (a) $\ket{10}$ or (b) $\ket{11}$. See section \ref{app:analyticerrors} for variable definitions, and Table \ref{tab:decay_params} for numeric values.}
    \label{fig:whokilledjfk}
\end{figure}

The probability of a blackbody decay that leaves the qubits in the configuration $QB$ (Fig.~\ref{fig:whokilledjfk}a) is given by the decay rate $\Gamma_B$ and the average population in the Rydberg state during the gate, $\alpha$:
\begin{equation}
    \Gamma_B \alpha t_g  = \Gamma_B \int_0^{t_g} \left|\psi_r(t)\right|^2 dt
\end{equation}
Similarly, the probability of a radiative decay to $QR$ is $\Gamma_R \alpha t_g$. For the LP gate, $\alpha \approx 0.532$.

The probability of the qubit decaying back to the computational space is $\Gamma_Q \alpha t_g$. We make two simplifying assumptions about this process. First, we set the decay probability to $\ket{00}$ and $\ket{01}$ to be equal, though in reality they are biased towards $\ket{01}$, which is more favorable. Second, we assume that the time spent in intermediate states is negligible compared to $t_g$, which is well-justified if $t_g > 100$ ns. After decaying to $\ket{00}$, the qubits will remain there for the rest of the gate. Decays to $\ket{01}$, however, result in re-excitation, resulting in $\ket{0r}$ population at the end of the gate, which is detected as a $QB$ configuration. We denote the fraction of decays to $\ket{01}$ that are re-excited as $R_{01}$, which we compute as a weighted average over the possible decay times:

\begin{equation}
    R_{01} = \frac{1}{t_g \alpha} \int_0^{t_g} \left|\psi_r(t)\right|^2 \left|\psi_r(t_g-t)\right|^2 dt \approx 0.700
\end{equation}

Here, $\left|\psi_r(t_g-t)\right|^2$ is the probability for an atom that has decayed at a time $t$ to be found in $\ket{r}$ at the end of the gate. To see why this is the case, consider the directly computed re-excitation probability:  $\left|\mel{r}{U(t,t_g)}{1}\right|^2$, where $U(t,t_g)$ is the propagator from time $t$ to $t_g$. Taking the complex conjugate inside the square modulus allows this to be rewritten as $\left|\mel{r}{U(t_g,t)}{1}\right|^2$, describing the evolution of $\ket{1}$ \emph{backwards} in time, from $t_g$ to $t$. Because the square modulus of the wavefunctions are clearly symmetric around the middle of the gate (Fig.~\ref{fig:gatesim_si}b), this can be replaced by $\left|\mel{r}{U(0,t_g-t)}{1}\right|^2 = \left|\psi_r(t_g-t)\right|^2$.

We can combine these results to arrive at the probability to end up in each subspace, having started in $\ket{01}$:
\begin{align}
\label{eq:p01_first}
    P(QR|01) &= \Gamma_R \alpha t_g \\
    P(QB|01) &= \Gamma_B \alpha t_g + (\Gamma_Q/2) \alpha t_g R_{01} \\
    P(QQ|01) &= 1 - P(QR|01) - P(QB|01)
\label{eq:p01_last}
\end{align}

\subsubsection{Initial state $\ket{11}$}

Now we consider the case that the qubits start in $\ket{11}$ (Fig.~\ref{fig:whokilledjfk}b). During the gate, with no errors, the state can be represented as:
\begin{equation}
\ket{\psi(t)} = \psi_{11}(t)\ket{11} + \psi_W(t)\ket{W} + \psi_{rr}(t) \ket{rr}
\end{equation}
where $\ket{W} = (\ket{1r} + \ket{r1})/\sqrt{2}$. We assume $|\psi_{rr}(t)|^2 \ll 1$ because of the Rydberg blockade, and neglect this component unless otherwise stated. The Rydberg excitation probability $|\psi_W(t)|^2$ is plotted in Fig.~\ref{fig:gatesim_si}b.

Proceeding as before, the probability of a blackbody decay to the subspace $QB \cup BQ$ depends on the average Rydberg population $\beta$:
\begin{equation}
    \Gamma_B \beta t_g = \Gamma_B \int_0^{t_g} \left|\psi_W(t)\right|^2 dt
\end{equation}
Similarly, the probability of a radiative decay to $QR \cup RQ$ is $\Gamma_R \beta t_g$. For the LP gate, $\beta \approx 0.467$. 

The qubits can also decay to back to the computational space $QQ$, with a total probability $\Gamma_Q \beta t_g$, and we assume that decays to $\ket{01}$ and $\ket{11}$ happen instantly with equal probability, as discussed in the preceding section. If the decay is to $\ket{11}$, then re-excitation can result in the configuration $QB \cup BQ$ at the end of the gate, with probability $R'_{11}$:
\begin{equation}
    R'_{11} = \frac{1}{t_g \beta} \int_0^{t_g} \left|\psi_W(t)\right|^2 \left|\psi_W(t_g-t)\right|^2 dt \approx 0.700
\end{equation}

If the decay is to $\ket{01}$, then re-excitation is also possible but with a different probability $R'_{11}$, given by the single-atom excitation trajectory $\psi_r$:
\begin{equation}
    R_{11} = \frac{1}{t_g \beta} \int_0^{t_g} \left|\psi_W(t)\right|^2 \left|\psi_r(t_g-t)\right|^2 dt \approx 0.640
\end{equation}

It is also possible that both atoms leave $Q$, resulting in the configurations $BR \cup RB$, $RR$ or $BB$. The configuration $BR \cup RB$ can be populated by an initial radiative decay to $QR \cup RQ$, followed by re-excitation of the qubit remaining in $Q$ (which is always in $\ket{1}$). The probability for this to occur is also $R_{11}$.

The configuration $RR$ can be populated by a second radiative decay after an initial decay to $QR \cup RQ$. The probability for this to occur, conditioned on the first radiative decay, is given by $\Gamma_R \beta' t_g$, where $\beta'$ is the average Rydberg population after the first decay:
\begin{equation}
\begin{split}
    \beta' &= \frac{1}{t_g \beta} \int_0^{t_g} dt \left|\psi_W(t)\right|^2 \frac{1}{t_g} \int_t^{t_g} dt' \left|\psi_r(t_g-t)\right|^2 \\
    &\approx 0.266
\end{split}
\end{equation}

Lastly, the configuration $BB$ can be populated in two ways: by re-excitation after an initial blackbody decay to $QB \cup BQ$, or decay from the doubly excited state $\ket{rr}$. The former is strongly suppressed by the blockade term $V_{rp}$ (in Eq. \ref{eq:ham}), while the latter is strongly suppressed by the blockade $V_{rr}$. The direct decay from $\ket{rr}$ occurs with probability:
\begin{equation}
    2\Gamma_B \beta'' t_g = 2 \Gamma_B \int_0^{t_g} \left|\psi_{rr}(t)\right|^2 dt
\end{equation}

with the average $\ket{rr}$ population $\beta'' \approx \beta \Omega^2/(2 V^2_{rr})$. Note that only a single decay is required, as the state $\ket{rp}$ results in the creation of two ions.

The probability for a pair of atoms that has already decayed to $QB \cup BQ$ to be re-excited is:
\begin{equation}
\begin{split}
    S &= \frac{1}{t_g \beta} \int_0^{t_g} \left|\psi_W(t)\right|^2 \left|\psi_{rp}(t_g-t)\right|^2 dt
\end{split}
\end{equation}

Here, $|\psi_{rp}(t)|^2 \approx \Omega^2/(2 V_{rp}^2)$ is the probability for the state $\ket{1p}$ to evolve into $\ket{rp}$ after a time $t$.

\begin{table}[]
    \centering
    \begin{tabular}{l|l}
         Term & Value  \\
         \hline
         $\alpha$ & 0.532 \\
         $R_{01}$ & 0.700 \\
         \hline
         $\beta$ & 0.467 \\
         $R_{11}$ & 0.640 \\
         $R'_{11}$ & 0.700 \\
         $\beta'$ & 0.266 \\
         $\beta''$ & $\beta \Omega^2/(2 V^2_{rr})$ \\
         $S$ & $\Omega^2/(2 V_{rp}^2)$ \\
    \end{tabular}
    \caption{Coefficients of the transition rates in Fig.~\ref{fig:whokilledjfk}, evaluated for the CZ gate from Ref. \cite{Levine2019}.}
    \label{tab:decay_params}
\end{table}

From these expressions, we can compute the probability to end up in different final states, starting in $\ket{11}$.

\begin{align}
\label{eq:p11_first}
    P(QR\cup RQ|11) &= \Gamma_R \beta t_g (1-R_{11}) \\
    \begin{split}
    P(QB\cup BQ|11) &= \Gamma_B \beta t_g (1-S) \\ &+ \Gamma_Q \beta t_g(R_{11} + R'_{11})/2
    \end{split}\\
    P(RB\cup BR|11) &= \Gamma_R \beta t_g R_{11} \\
    P(RR | 11) &= (\Gamma_R t_g)^2 \beta \beta' \\
    P(BB | 11) &= \Gamma_B \beta t_g S + 2 \Gamma_B \beta'' t_g
\label{eq:p11_last}
\end{align}

\subsubsection{Summary}

We can combine the analytic estimates above in Eqs. \eqref{eq:p01_first}-\eqref{eq:p01_last} and Eqs. \eqref{eq:p11_first}-\eqref{eq:p11_last} to obtain a total probability of each error channel. Given an initial state with probability $\{P_{00},P_{01},P_{11}\}$ to be in $\{\ket{00},\ket{01} or \ket{10},\ket{11}\}$, the probability of each error channel is:

\begin{align}
\label{eq:finalprobfirst}
    P_{QR} &= P_{01} \Gamma_R \alpha t_g + P_{11} \Gamma_R \beta t_g (1-R_{11}) \\
    \begin{split}
    P_{QB} &= P_{01} \Gamma_B \alpha t_g \\
    &+ P_{11} \left[\Gamma_B \beta t_g (1-S) + \Gamma_Q \beta t_g (R_{11}+R'_{11})/2\right]
    \end{split} \\
    P_{RB} &= P_{11} \Gamma_R \beta t_g R_{11} \\
    P_{RR} &= P_{11} (\Gamma_R t_g)^2 \beta \beta' \\
    P_{BB} &= P_{11} \left[2 \Gamma_B \beta'' t_g + \Gamma_B \beta t_g S\right]
\label{eq:finalproblast}
\end{align}

The total erasure probability $p_e$ is given by the sum of the first five terms. The probability of an undetectable leakage error is $p_f = P_{BB}$.

The first three errors scale as $t_g$; correspondingly, the probability of these events goes as $\Gamma t_g$, and are the dominant error mechanism for the gate. The fourth expression, $P_{RR}$, decreases as $(\Gamma t_g)^2$. The final error probability $P_{BB}$, scales as $\Gamma t_g \Omega^2/(2 V^2) \approx \Gamma / (t_g V^2)$ (here, $V$ is the smaller of $V_{rr},V_{rp}$, which is typically $V_{rr}$. While this error probability decreases with $\Gamma$, it increases as $t_g$ decreases, as the larger $\Omega$ begins to overpower the blockade. As noted in the main text, the error $BB$ is special because it cannot be readily detected and results in atom loss. However, excitation of $\ket{rr}$ causes other, coherent errors in the gate as well. Therefore, maintaining high fidelity gate operation even in the absence of spontaneous decay requires $\Omega/V > 20$ \cite{Levine2019}. Since $P_{BB}/P_{QB} \approx \Omega^2/(2 V^2)$, it seems that the probability of $BB$ events will generally be smaller than the probability of undetected $QB$ events, given the detection fidelity discussed in section \ref{app:detection}.

A final source of error is the non-Hermitian no-jump evolution that arises under the monitoring realized by the erasure detection \cite{plenio1998}. Since erasure errors do not occur from the state $\ket{00}$, and are approximately equally likely from the remaining computational states, the absence of an erasure detection reveals that the atoms are more likely to be in $\ket{00}$. The impact on the average gate infidelity is approximately $(p_e/4)^2$, which is not a significant contribution when $p_e \ll 16(1-R_e) \approx 0.32$.

\subsection{Comparison to numerical simulations}

For comparison, we also perform a master equation simulation of the full two-atom model. We consider the error probabilities as a function of the gate duration, $t_g$, which depends on the Rabi frequency as $t_g \approx 8.586/\Omega$. The gate error depends primarily on the dimensionless quantity $\Gamma t_g$, but is also sensitive to the blockade strength (in the high-fidelity regime), which we express in dimensionless units as $V_{rr}/\Gamma$. For simplicity, we set $V_{rp} = V_{rr}$, though in reality, $V_{rp}$ is larger because it is a first-order process.

For the $n=75$ $^3S_1$ state in $^{171}$Yb, we assume a Rydberg lifetime $\tau = 1/\Gamma = 100$ $\mu$s, and $V=2\pi \times 1.3$ GHz, based on previous measurements in $^{174}$Yb \cite{Wilson2019,burgers2021}, giving $V/\Gamma = 10^6$. The achievable value of $t_g$ depends on the details of the experimental setup and excitation laser. However, we note that $\Omega = 2\pi \times 5.5$ MHz has been demonstrated for this state (starting from $^3P_1$) with very modest laser power \cite{burgers2021}, which would yield $t_g \approx 250$ ns and $\Gamma t_g \approx 2 \times 10^{-3}$.

In Fig.~\ref{fig:gatesim_si}c, the predictions of Eqs. \eqref{eq:finalprobfirst}-\eqref{eq:finalproblast} are shown along with a master equation simulation of the two-atom model. The numerical simulation and the analytic model are in excellent agreement.

\section{Erasure conversion for other errors}
\label{app:othererrors}

While we have so far focused on two-qubit gate errors, as they are dominant and most problematic, the metastable state qubit encoding in $^{171}$Yb should also allow erasure conversion for other errors. In this section, we briefly sketch these ideas, leaving a detailed analysis for future work.

First, any spontaneous decay or photon scattering occurring on idle qubits in the ${}^{3}$P$_{0}$ level is an erasure error with very high probability. Spontaneous decay to $^1$S$_0$ is always detectable. Raman and Rayleigh scattering from the optical tweezer have a vanishing probability of creating Pauli errors in the qubit subspace as long as the tweezer detuning is large compared to the hyperfine splitting in other excited states \cite{dorscher2018}. It can shorten the lifetime of the qubit level by Raman scattering to other $^3$P$_J$ states, but these decay or are repumped to $^1$S$_0$, and detected as erasures.

The same logic can be applied to single-qubit gates performed using Raman transitions via the $6s7s$ $^3$S$_1$ state, as long as the detuning is large compared to the hyperfine splitting in that state \cite{Ozeri2007}. If single-qubit gates are performed through the Rydberg state, then the analysis is the same as that of the two-qubit gate.

Lastly, we note that a significant source of error in current neutral atom gates is technical noise, either from Doppler shifts or frequency and intensity fluctuations of the driving laser. While this source of error is not fundamental, it is a significant practical nuisance. Noise that is slow compared to the duration of a gate, which is often the case for Doppler shifts and intensity noise, can be cancelled using composite pulse sequences \cite{wolfowicz2016} or other robust control techniques \cite{wenzheng2021}. Unfortunately, this typically results in a longer total gate duration, increasing the Rydberg decay probability. However, this trade-off may be more advantageous with erasure conversion.

\section{Impact of errors in initialization, measurement and single-qubit gates}

\begin{figure}
    \centering
    \includegraphics{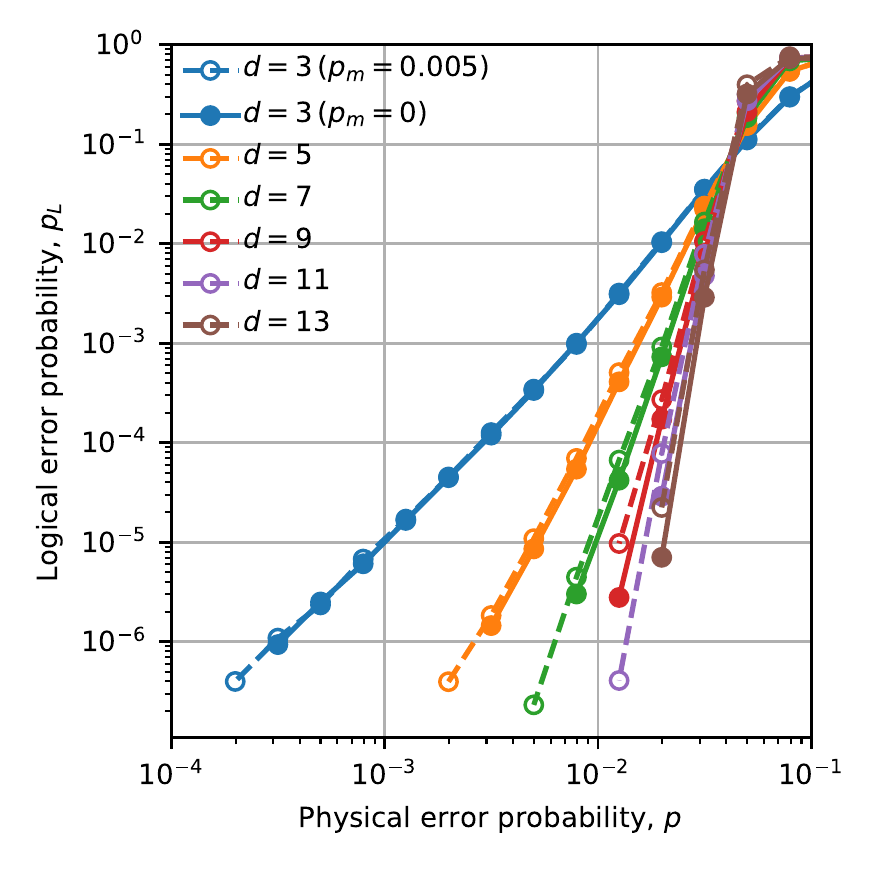}
    \caption{Logical error performance with errors on ancilla qubit initialization and measurement with probability $p_m=0$ (solid lines, filled circles) and $p_m=0.005$ (dashed lines, open circles).}
    \label{fig:pm_si}
\end{figure}

In the simulations in Figs, 3 and 4, we assume native operations to initialize and measure the ancillas in the Hadamard basis, and native CNOT and CZ gates, such that no single-qubit gates are required. The impact of single-qubit gate errors can be estimated by considering an alternative stabilizer measurement circuit with ancilla initialization and measurement in the Z basis, and only CZ gates. This requires the insertion of four $H$ gates, which can each be associated with one of the four two-qubit gates. Therefore, a pessimistic assumption is to treat an error in the $H$ as an error in the two-qubit gate, which would increase the two-qubit gate error probabilities to $p_p + p^{(1)}_p$ and $p_e + p^{(1)}_e$, where $p^{(1)}_e/(p^{(1)}_p+p^{(1)}_e) = R^{(1)}_e$ is the erasure fraction of the single qubit gate. If $R^{(1)}_e=0$, then $R_e$ is reduced by a factor $1/(1+p^{(1)}/p)$, which means that $R_e$ is not significantly affected if $p^{(1)}/p < 1-R_e$. This is not an unreasonable assumption for $R_e = 0.98$. However, as discussed above, it is also possible to extend erasure detection to single-qubit gates, which would further relax this requirement.

Additionally, we consider the role of imperfect ancilla initialization and measurement. In the simulation, this is represented by inserting Pauli errors before or after perfect operations with probability $p_m$. Note that $p_m=0$ in Figs.~\ref{fig:pl}, \ref{fig:dist}. Here we attempt to quantify the impact of realistic initialization and measurement errors in two ways. First, we consider a fixed value of $p_m$. For $p_m=0.001$, the threshold two-qubit gate error for $R_e=0.98$ is indistinguishable from its value when $p_m=0$. If $p_m = 0.005$, we find that the threshold is slightly reduced to $p_{th} = 3.80(2)\%$, but the general behavior, even far below the threshold, is unchanged (Fig.~\ref{fig:pm_si}). Second, we study the case that the initialization and measurement errors have the same probability as two-qubit gate errors, $p_m=p$. In this case, we find the threshold decreases to 2.85(1)\%.

\section{Impact of erasure conversion on operation speed}

In this section we consider how the operations required for erasure conversion may affect the overall computation speed of a neutral atom quantum computer. A single round of stabilizer measurements for a surface code with distance $d$ requires of order $N=4d^2$ two-qubit gates, each of which takes a duration $t_g < 1\,u$s. Gates that are sufficiently remote can be implemented in parallel \cite{Levine2019,Graham2019}, and we estimate that in the limit of a large array, a fraction $f_p=1/10$ of the gates can be applied in each cycle. Therefore, the total time required to apply the gates is $t_g/f_p \approx 10\,\mu$s.

The erasure detection step must occur after each set of parallel gates, and takes a time $t_e \approx 10\,\mu$s (as discussed in section \ref{app:detection}). This increases the cycle time to $(t_g+t_e)/f_p \approx 100\,\mu$s.

Atom replacement can be deferred until after the stabilizer measurement: once an erasure error has occurred, subsequent gates involving the affected atoms can simply be skipped. The time to move tweezers is $t_r$, which is several hundred microseconds in recent experiments, \cite{bluvstein2021}. All necessary replacements can be performed in parallel. 

Lastly, the ancilla qubits need to be measured to extract the syndrome values, and we denote the time for this operation as $t_m$. To enable the atoms to be re-used, this measurement should not result in the loss of atoms, which limits the scattering rate and results in $t_m \gtrsim 20$ ms \cite{Saskin2019,ma2021}.

Therefore, the total duration of a cycle is $(t_g+t_e)/f_p + t_r + t_m$. This is dominated by $t_m$, and therefore, the erasure conversion protocol will not significantly affect the total repetition time unless $t_m$ is reduced by about two orders of magnitude \cite{xu2021}.

\end{document}